\documentclass[aps,prl,showpacs,a4paper,twocolumn]{revtex4} 
\usepackage{amsmath}
\usepackage{amsfonts}
\usepackage{bm}
\usepackage{dcolumn}
\usepackage{graphicx}

\begin{document}

\author{Beno{\^\i}t Gr{\'e}maud}
\email{Benoit.Gremaud@spectro.jussieu.fr}
\affiliation{Laboratoire Kastler Brossel, Universit{\'e} Pierre et
Marie Curie, Case 74,\\
4, place Jussieu, 75252 Paris Cedex 05, France}
\title{Semi-classical analysis of real
  atomic spectra beyond Gutzwiller's approximation}
\date{\today}

\begin{abstract}
  Real atomic systems, like the hydrogen atom in a magnetic field or the
  helium atom, whose classical dynamics are chaotic, generally present
  both discrete and continuous symmetries. In this letter, we explain
  how these properties must be taken into account in order to obtain
  the proper (i.e. symmetry projected) $\hbar$ expansion of
  semiclassical expressions like the Gutzwiller trace formula. In the
  case of the hydrogen atom in a magnetic field, we shed light on the excellent
  agreement between present theory and exact quantum results.
\end{abstract} 
\maketitle

In the studies of the quantum properties of systems whose classical
counterparts depict chaotic behavior, semi-classical formulas are
essential links between the two worlds, emphasized by Gutzwiller's
work~\cite{Gutzwiller90}. More specifically, starting from Feynman's path
formulation of quantum mechanics, he has been able to express the 
quantum density of states as a sum over all (isolated) periodic orbits of
the classical dynamics. This
formula, and extensions of it, have been widely used to understand and obtain
properties of the energy levels of many classically chaotic
systems, among which the
hydrogen atom in a magnetic field~\cite{FW89,Houches89}, the helium
atom~\cite{Ezra91,GR93,GG98} or
billiards~\cite{Bu79,CE89,GA93,GAB95}.

At the same time, because the trace formula (and its variations) as
derived by Gutzwiller only 
contained the leading term of the asymptotic expansion of the quantum level
density, the systematic expansion of the semiclassical propagator in
powers of $\hbar$ has been the purpose of several
studies~\cite{GA93,GAB95,VR96,WMW02}, but which focused on billiards, for
which both classical and quantum properties are 
easier to calculate.

In a recent paper~\cite{pre65G}, general equations for efficient
computation of $\hbar$ corrections in semi-classical formulas for a
chaotic system with smooth dynamics were presented, together with
explicit calculations for the hydrogen atom in a magnetic field.
However, only the two-dimensional case was considered, because for the
three-dimensional (3D) case, discrete symmetries and centrifugal terms
had to be taken into account. Actually, this situation occurs in
almost all real atomic systems depicting a chaotic behavior
(molecules, two electron atoms...), for which experimental data
involve levels having well defined parity, total angular momentum and,
if relevant, exchange between particles.  In particular,
semi-classical estimations of experimental signals like
photoionization cross-sections are calculated with closed orbits with
vanishing total angular momentum, whereas they usually involve $P$
($L=1$) quantum states, whose positions in energy are shifted with
respect to $S$ ($L=0$) states.  Furthermore, in recent years, the
development of the harmonic inversion method makes it possible to
extract the relevant quantities (position of peaks, complex
amplitudes) from both theoretical and experimental data with a much
higher accuracy than with the conventional Fourier
transform~\cite{M99}. In particular, it becomes possible to measure the
deviation of the exact quantum results from the semi-classical leading
order predictions.  Thus, a detailed semi-classical analysis of
experimental results, beyond the leading order in $\hbar$, requires
the understanding and the calculation of corrections due to both the
discrete symmetries and centrifugal terms.
In addition, we would like to stress that even if the present analysis
is made with the density of states, it can
also be made with the Quantum Green function, which leads to
expressions and numerical computations of the first order $\hbar$
corrections for physical quantities like the photo-ionization
cross-section~\cite{B89,GD92}, which could either be compared to
available experimental data~\cite{Holle88,Main94}, or become a
starting point for refined experimental tests of the quantum-classical
correspondence in the chaotic regime. 

$\hbar$ corrections and discrete symmetries have
already been discussed, but only for
billiards~\cite{GA93,GAB95,WMW02}, whereas in the case of systems with smooth
dynamics a detailed study is still lacking. 
Also, centrifugal terms and/or rotational
symmetries have been considered by many authors, but
either in the case of integrable systems~\cite{schulman,KL75}, or for
values of the angular momentum comparable to the action of classical
orbits~\cite{Gutzwiller90,praCL44,jpaCL25}. From this point of view, 
the present study, which focuses on fixed values of the quantum
angular momentum and the effect of the centrifugal terms on $\hbar$
corrections for systems with smooth chaotic dynamics, goes beyond the
preceding considerations. 
More precisely, in this letter, we explain
how to take into account both discrete
symmetries and centrifugal terms
 in order to obtain a full semi-classical description of
the first order $\hbar$ corrections for the 3D hydrogen atom in a
magnetic field.

At first, 
in the case of a chaotic system, whose Hamiltonian $H=\mathbf{p}^2/2+V(\mathbf{q})$ is invariant under a group $\mathcal{S}$ of discrete
transformations $\sigma$, the leading order of
semi-classical approximation for the trace of the Green function
$G(E)=1/(E-H)$,
restricted to the $m^{\mathrm{th}}$ irreducible representation is 
given by~\cite{pra40R}:
\begin{equation}
\label{symtrace}
g_m^{\mathrm{sc}}(E)=\frac{d_m}{i\hbar}\sum_l\frac{T_l}{|K_l|}\sum_n
\chi_m(\sigma_l^n)g_{(l,n)}^{(0)}(E)
\end{equation}
with
\begin{equation}
g_{(l,n)}^{(0)}(E)=\frac{1}{|\det{(A^n_l-\openone)}|^{1/2}}
\exp{\left[\frac{i}{\hbar}nS_l-in\mu_l\frac{\pi}{2}\right]}
\end{equation}
where the $l$ sum is taken over all primitive (isolated) orbits which
become periodic through the symmetry operation $\sigma_l$ 
(i.e. final position (resp. velocity) is 
mapped back to initial position (resp. velocity) by
$\sigma_l$). $\chi_m(\sigma_l^n)$ is the character 
of $\sigma_l^n$ in the $m^{\text{th}}$ irreducible
representation of dimension $d_m$. $S_l$ is the action of the orbit
$l$, $\mu_l$ the 
Maslov index, $T_l$ the ``period'', $A^n_l$
represents the Poincar{\'e} surface-of-section map linearized around the
orbit and $K_l$ is
the subgroup of $\mathcal{S}$ leaving each point of the orbit $l$
invariant.
Adding first order $\hbar$ corrections, the preceding
equation~\eqref{symtrace} becomes~:
\begin{equation}
\label{symtrace1}
g_m^{\mathrm{sc}}(E)=\frac{d_m}{i\hbar}\sum_l\frac{T_l}{|K_l|}
\sum_n\chi_m(\sigma_l^n)
g_{(l,n)}^{(0)}(E)\biggl(1+i\hbar\mathcal{C}_{(l,n)}^{\mathrm{tr}}\biggr)
\end{equation}

$\mathcal{C}_{l,n}^{\mathrm{tr}}$ 
can be derived by a detailed analysis of the
stationary phase approximations starting from the Feynman path integral,
following the same steps as in Ref.~\cite{GAB95,pre65G} and reads as
follows~:
\begin{equation}
\mathcal{C}_{l,n}^{\mathrm{tr}}=C_{l,n}^{T\rightarrow E}+
\frac{1}{nT_{l}}\int_0^{nT_l}\!dt_0\,
C_{l,n}(t_0)
\end{equation}
where $C_{l,n}^{T\rightarrow E}$ arises from the time to energy domain
transformation. $C_{l,n}(t_0)$ (see Ref.~\cite{pre65G} for the
expressions) involves the classical Green 
functions $\mathcal{G}_{l,n}(t,t')$, \textit{i.e.} the solutions of the
equations controlling the linear stability around the classical
trajectory $\mathbf{q}^{\mathrm{cl}}_{l,n}(t)$~:
\begin{equation}
\biggl(-\frac{\mathrm{d}^2}{\mathrm{d}t^2}\openone-
\frac{\partial^2V}{\partial\mathbf{q}\partial\mathbf{q}}
\left[\mathbf{q}^{\mathrm{cl}}_{l,n}(t)\right]\biggr)
\mathcal{G}_{l,n}(t,t')=\openone\,\delta(t-t').
\end{equation}
The fact
that the orbits are periodic after the symmetry transformation 
$\sigma_l^n$ determines the boundary conditions that the classical Green 
functions $\mathcal{G}_{l,n}(t,t')$ must fulfill, namely~:
\begin{equation}\left\{
\begin{aligned}
\sigma_l^{-n}\mathcal{G}_{l,n}(nT_l,t')&=\mathcal{G}_{l,n}(0,t') \\
\mathcal{P}_{t_0}\mathcal{G}_{l,n}(0,t')&=0 \\
\mathcal{Q}_{t_0}\sigma_l^{-n}\dot{\mathcal{G}}_{l,n}(nT_l,t') &=
\mathcal{Q}_{t_0}\dot{\mathcal{G}}_{l,n}(0,t')
\end{aligned}
\right. \qquad \forall t'\in[0,nT_l]
\end{equation}
where $\mathcal{P}_{t_0}$ is the projector along the ``periodic'' orbit at
the position depicted by  time $t_0$ and
$\mathcal{Q}_{t_0}=\openone-\mathcal{P}_{t_0}$. 
Of course, for $\sigma_l=\openone$, one recovers the boundary conditions given
in Ref.~\cite{pre65G}. Finally, all technical steps of
Ref.~\cite{pre65G} leading to efficient  
computation  of $\mathcal{G}_{l,n}(t,t')$ and $\hbar$ corrections,
that is, solutions of  sets of
first order differential equations, can easily be adapted
to take into account these modified boundary conditions.

\begin{figure}[ht]
\includegraphics[angle=270,width=8cm]{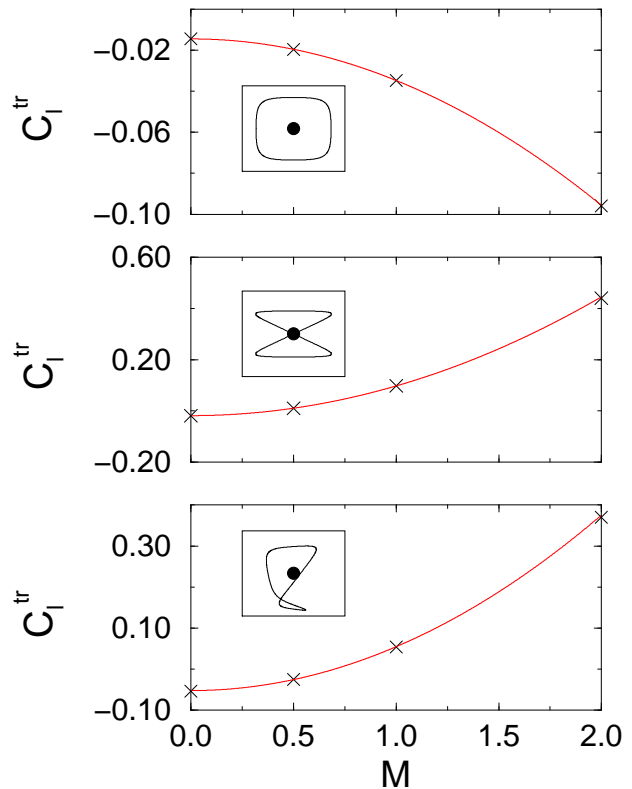}
\caption{\label{angfig}First order $\hbar$ correction to the
  semi-classical approximation of the trace of the quantum Green
  function for the hydrogen atom in a magnetic field for different
  values of the magnetic number $M$, $M=1/2$ corresponding to the 2D
  case~\cite{pre65G}. Crosses depict the values 
  extracted from the exact quantum function using harmonic inversion,
  whereas the solid line corresponds to the classical results given by
  Eq.~\eqref{c1smooth}. For the three different periodic
  orbits, whose trajectories in the $(u,v)$ plane are plotted (the
  nucleus being depicted by the black dot), the agreement is
  excellent, thus emphasizing the validity of Eq.~\eqref{smoothu} and 
Eq.~\eqref{c1smooth}.}
\end{figure}

As a numerical example, we have considered the 2D
hydrogen atom in a magnetic field,
 at scaled energy
$\epsilon=-0.1$~\cite{FW89}. More precisely, we have computed the
trace of the quantum Green function, using roughly 8000 states
belonging to the $EEE$ representation~\cite{MWR89} of the group
$D_4$, corresponding to effective $1/\hbar$ values ranging from $0$ to
$124$ (See 
Ref.~\cite{pre65G} for further details). In
that case, the periodic orbit 
$\overline{1234}$~\cite{EW90,H95} (see inset of the top of
Fig.~\ref{angfig} for the trajectory in semi-parabolic coordinates),
being
(globally) invariant under a rotation of angle $\pi/2$,  gives rise to
contributions in 
the semi-classical approximation of the trace at all multiples of
$S_{\overline{1234}}/4$. In the same way, the periodic orbit
$\overline{1243}$ (see middle inset of Fig.~\ref{angfig}) being
invariant under a rotation of angle $\pi$, 
contributions are present at all multiples of
$S_{\overline{1243}}/2$. For both these orbits, 
table~\ref{tabpernum} displays the 
comparison of the present theoretical calculation and the numerical
coefficient $\mathcal{C}_{l,n}^{HI}$,
extracted from the exact quantum Green function, using harmonic
inversion~\cite{M99,pre65G}. As one can notice, the agreement is
excellent for the amplitudes and rather good for the phases, which is
the usual behavior of harmonic inversion. Furthermore, the same agreement
has also been found for the other representations, thus emphasizing 
the present approach for the calculation of the first order $\hbar$
corrections when taking into account discrete symmetries. 

\begin{table}[ht]
\caption{\label{tabpernum}Numerical comparison between the
  theoretical $\hbar$ corrections $\mathcal{C}_l^{\mathrm{tr}}$ for
  the trace of the quantum
  Green function, restricted to the $EEE$ representation,  of the 2D
  hydrogen atom in a magnetic field and the numerical
  coefficients $\mathcal{C}_l^{HI}$ extracted from exact quantum function  
using harmonic inversion. The
agreement is excellent for  
the amplitudes and rather good on the phases, thus emphasizing  the
validity of the present approach.} 
\begin{ruledtabular}
\begin{tabular}{l@{}l@{}l@{}l@{}r}
\multicolumn{1}{c}{Code} & 
\multicolumn{1}{c}{$\mathcal{C}_l^{\mathrm{tr}}$} & 
\multicolumn{1}{c}{$|\mathcal{C}_l^{HI}|$} & 
\multicolumn{1}{c}{Rel. error} &
\multicolumn{1}{c}{$\arg{\mathcal{C}_l^{HI}}$} \\
\hline
$\frac{1}{4}\overline{1234}$ & $-$0.094\,430 & 0.09445 & $\approx2\times10^{-4}$ 
& 0.9996$\times\pi$ \\
$\frac{1}{2}\overline{1234}$ & $-$0.361\,689 & 0.3611 & $\approx2\times10^{-3}$ 
& 0.996$\times\pi$ \\
$\frac{3}{4}\overline{1234}$ & $-$0.400\,555 & 0.3992 & $\approx3\times10^{-3}$ 
& 1.005$\times\pi$ \\
$\frac{1}{2}\overline{1243}$ & $\phantom{-}$0.049\,339 & 0.0493   & $\approx8\times10^{-4}$ 
& -0.075$\times\pi$    
\end{tabular}
\end{ruledtabular}
\end{table}

Contrary to the preceding, calculating first order $\hbar$ corrections
due to centrifugal terms is more 
complicated and is best explained in the case of the 3D hydrogen atom
in a magnetic field. The regularized Hamiltonian in semiparabolic
coordinates, for fixed value $M$ of the projection of the angular
momentum along the field axis, is given by~\cite{FW89}~:
\begin{equation}
\begin{aligned}
  H&=-\frac{\hbar^2}{2}
\left(\frac{\partial^2}{\partial u^2}+
\frac{\partial^2}{\partial v^2}
+(\frac{1}{4}-|M|^2)\left[\frac{1}{u^2}+\frac{1}{v^2}\right]\right)\\
&\qquad\qquad\qquad-\epsilon(u^2+v^2)+\frac{1}{8}u^2v^2(u^2+v^2) \\
&=H_0+\frac{\hbar^2}{2}\left(|M|^2-\frac{1}{4}\right)
U(u,v).
\end{aligned}
\end{equation}
$H_0$ is then the Hamiltonian of the 2D hydrogen atom in a magnetic field.
If $U(u,v)$ was regular, then the additional first order $\hbar$
correction for the orbit $l$ would simply be~:
\begin{equation}
\label{corr}
-\frac{1}{2}\left(|M|^2-\frac{1}{4}\right)\int_0^{T_l}\!dt\,
U(u_l(t),v_l(t)).
\end{equation}
One must mention that in this case, the Langer
transformation~\cite{L37} of the coordinates
$(u,v)\rightarrow(\exp{(-x)},\exp{(-y)})$ gives rise to a 
Hamiltonian which does not separate into kinetic and potential
energies and for which no expressions for $\hbar$ corrections are
available.

On the other hand, the fact that $U(u,v)$ is singular imposes
boundary conditions on 
both classical and quantum dynamics. The
classical trajectories have to make (smooth) bounces near $u=0$ and
$v=0$ and for vanishing values of $\hbar$, we expect
the trajectories of $H$ to be those of $H_0$, but  mapped onto
the reduced phase space $(u>0,v>0)$, i.e. making hard bounces on the
$(u,v)$ axis. From the quantum point of view,
depending on the parity of $M$, only  wavefunctions belonging to given
representations of $D_4$ are allowed. Thus, first
order $\hbar$ corrections 
due to the singular part of the potential $U$, are given by the
preceding considerations on the symmetries, whereas remaining
corrections are given by Eq.~\eqref{corr}, where $U$ has to be
replaced by a smooth counterpart, namely~:
\begin{equation}
\label{smoothu}
\tilde{U}=\lim_{\epsilon\rightarrow 0^+}
\frac{1}{2}\left(\frac{1}{(u+i\epsilon)^2}+\frac{1}{(u-i\epsilon)^2}
+\frac{1}{(v+i\epsilon)^2}+\frac{1}{(v-i\epsilon)^2}\right).
\end{equation}
Actually, one can show that the preceding equation gives the right
answers for $\hbar$ expansion of the propagator of the free particle
(up to $\hbar^3$)
and the harmonic oscillator (up to $\hbar^2$), for which
analytical expressions for classical trajectories, 
classical Green
functions and quantum propagators exist (higher orders have not been
checked yet).  
However, even if a detailed analysis of the derivation of the trace
formula in presence of centrifugal terms seems to show that the preceding
approach works in general cases, rigorous proof of
Eq.~\eqref{smoothu} is lacking.

Nevertheless, in the case of the 3D hydrogen atom in a magnetic field, we
have compared the first order $\hbar$ corrections, for different
periodic orbits and for different values of the magnetic number $M$,
with the present prediction, namely~:
\begin{equation}
\label{c1smooth}
C_l^{\mathrm{tr}}(M)=C_l^{\mathrm{tr}}(2D)-\frac{1}{8}\left(4|M|^2-1\right)\int_0^{T_l}\!dt\,
\tilde{U}(u_l(t),v_l(t)).
\end{equation}
The results are displayed in Fig.~\ref{angfig} for $M=0,1,2$ and for
three different 
orbits, namely $\overline{1234}$, $\overline{1243}$ and
$\overline{12343}$, whose trajectories in the $(u,v)$ plane are
plotted.  The solid line is the theoretical result given by
Eq.~\eqref{c1smooth}, whereas the crosses are the values extracted
from the trace of the exact quantum Green function, using harmonic
inversion (for scaled
energy $\epsilon=-0.1$, roughly 8000 effective $1/\hbar$ values ranging
from $0$ to $124$).   
As one can notice the agreement is excellent, thus giving strong
support for the validity of Eq.~\eqref{smoothu} and 
Eq.~\eqref{c1smooth}. Furthermore, the simplicity of the replacement
$\tilde{U}$ may serve as a guideline for a rigorous treatment of the
$\hbar$ corrections arising from the centrifugal terms. In particular,
the calculation of higher orders involves products of the derivatives
of these centrifugal terms and those of the potential $V_0$, giving
rise to non-trivial mixing between
 centrifugal and standard $\hbar$ corrections. 

In conclusion, we have presented a semi-classical analysis, beyond the
usual Gutzwiller approximation, including first order $\hbar$ corrections, of
the quantum properties of real chaotic systems. 
More specifically, we have explained the additional corrections
arising when taking into account both discrete symmetries and
centrifugal terms. In the case of the (3D) hydrogen in a magnetic field,
the agreement between the theory and the numerical
data extracted from exact quantum results is excellent, emphasizing
the validity of the analysis, especially of equations~\eqref{smoothu} and
\eqref{c1smooth}. 

Finally, since we
know how to compute the $\hbar$ corrections, it would be very interesting
to work the other way round, that is, to perform the semi-classical
quantization, thus getting  $\hbar$ corrections in the
semi-classical estimations of the quantum quantities, like the
eigenenergies. Of course, this represents a more considerable amount of
work, since the $C_{l,n}^{\mathrm{tr}}$ coefficients must be computed
for all relevant orbits and then included in standard semi-classical
quantization schemes, like the cycle expansion~\cite{EC89,GR93,VR96}.

\begin{acknowledgments}
The author thanks D.~Delande for his kind support during this work.
Laboratoire Kastler Brossel is laboratoire de l'Universit{\'e} Pierre et Marie
Curie et de l'Ecole Normale Sup{\'e}rieure, unit{\'e} mixte de
recherche 8552 du CNRS.
\end{acknowledgments}


\begin{thebibliography}{99} 
\bibitem{Gutzwiller90} \textit{Chaos in Classical and Quantum
    Mechanics}, M.C.~Gutzwiller (Springer, New~York, 1990).


\bibitem{FW89} H.~Friedrich and D.~Wintgen,
  Phys.~Rep. \textbf{183}, 37 (1989).

\bibitem{Houches89} D.~Delande, in \textit{Chaos and Quantum Physics},
edited by M.-J.~Giannoni, A.~Voros, and J.~Zinn-Justin, Les Houches
Summer School, Session LII (North-Holland, Amsterdam, 1991).

\bibitem{Ezra91} G.S.~Ezra, K.~Richter, G.~Tanner, and D.~Wintgen,
  J.~Phys.~B \textbf{24}, L413 (1991).

 \bibitem{GR93} P.~Gaspard and S.A.~Rice. Phys.~Rev.~A \textbf{48},
54 (1993).

\bibitem{GG98} B.~Gr{\'e}maud and P.~Gaspard, J.~Phys.~B \textbf{31},
  1671 (1998).

\bibitem{Bu79} L.A.~Bunimovitch, Commun.~Math.~Phys. \textbf{65}, 295
  (1979).

\bibitem{CE89} P.~Cvitanovi\'c and B.~Eckhardt,
  Phys.~Rev.~Lett. \textbf{63}, 823 (1989).

\bibitem{GA93} P.~Gaspard and D.~Alonso, Phys.~Rev.~A
  \textbf{47}, R3468 (1993).

\bibitem{GAB95} P.~Gaspard, D.~Alonso, and I.~Burghardt,
Adv.~Chem.~Phys. \textbf{XC} 105 (1995).

\bibitem{VR96} G.~Vattay and P.~E.~Rosenqvist,
  Phys.~Rev.~Lett. \textbf{76}, 335 (1996).

\bibitem{WMW02} K.~Weibert, J.~Main, G.~Wunner, Eur.~Phys.~J.~D 
\textbf{19},  379 (2002).

\bibitem{pre65G} B.~Gr{\'e}maud, Phys.~Rev.~E \textbf{65},  056207
  (2002).

\bibitem{M99} J.~Main, Phys.~Rep. \textbf{316}, 233 (1999).

\bibitem{B89} E.~P.~Bogomolny, Zh. Eksp. Teor. Fiz. \textbf{96},
  487 (1989) [Sov. Phys. JETP \textbf{69}, 275 (1989)].


\bibitem{GD92} J.~Gao and J.~B.~Delos, Phys. Rev. A \textbf{46}, 1455
  (1992).

\bibitem{Holle88} A.~Holle, J.~Main, G.~Wiebusch, H.~Rottke,
  and K.~H.~Welge, Phys. Rev. Lett. \textbf{61}, 161 (1988).

\bibitem{Main94} J.~Main, G.~Wiebusch, K.~Welge, J.~Shaw, and
  J.~B.~Delos, Phys. Rev. A \textbf{49}, 847 (1994).

\bibitem{schulman} \textit{Techniques and Applications of Path
    Integration}, L.S.~Schulmann, (J.~Wiley, New-York, 1981).

\bibitem{KL75} K.C.~Khandekar and S.V.~Lawande,
  J.~Math.~Phys. \textbf{16}, 384 (1975).

\bibitem{praCL44} S.C.~Creagh and R.~G.~Littlejohn, Phys.~Rev.~A
  \textbf{44}, 836 (1991)

\bibitem{jpaCL25} S.C.~Creagh and R.~G.~Littlejohn, J.~Phys.~A
  \textbf{25}, 1643 (1992)



\bibitem{pra40R} J.~M.~Robbins, Phys.~Rev.~A \textbf{40}, 2128 (1989).



\bibitem{MWR89} C.~C.~Martens, R.~L.~Waterland, and W.~P.~Reinhardt,
  J.~Chem.~Phys. \textbf{90}, 2328 (1989).

\bibitem{EW90}  B.~Eckhardt and D.~Wintgen, J.~Phys.~B \textbf{23},
  355 (1990).

\bibitem{H95} K.~T.~Hansen, Phys.~Rev.~E \textbf{51}, 1838 (1995).



\bibitem{L37} R.E.~Langer, Phys.~Rev. \textbf{51}, 669 (1937)




\bibitem{EC89} P.~Cvitanovi\'c  and  B.~Eckhardt,
  Phys. Rev. Lett. \textbf{63}, 823 (1989) 

\end{thebibliography}
\end{document}